\documentclass[aps,prl,preprint,showpacs,showkeys,superscriptaddress]{revtex4-1}
\usepackage[pdftex]{graphicx} %[dvips]
\usepackage{natbib}
\usepackage{booktabs}
\usepackage{color}
\usepackage{url}
\newcommand{\bv}[1]{\mathbf{#1}}

\newcommand{\CRO}{Ca$_{2}$RuO$_{4}$}
\newcommand{\SRO}{Sr$_{2}$RuO$_{4}$}
 
\newcommand{\LCRO}{Ca$_{2-x}$La$_{x}$RuO$_{4}$} 
\newcommand{\CPRO}{Ca$_{2-x}$Pr$_{x}$RuO$_{4}$} 
\newcommand{\xy}{$d_{xy}$}
\newcommand{\xz}{$d_{xz}$}
\newcommand{\yz}{$d_{yz}$}
\newcommand{\LPbca}{\mbox{\textit{L-Pbca}}}
\newcommand{\SPbca}{\mbox{\textit{S-Pbca}}}

\begin{document}
\title{\textit{In-situ} strain-tuning of the metal-insulator-transition of Ca$_{2}$RuO$_{4}$ in angle-resolved photoemission experiments}

\author{S. Ricc\`{o}}
\affiliation{Department of Quantum Matter Physics, University of Geneva, 24 Quai Ernest-Ansermet, 1211 Geneva 4, Switzerland}
\author{M. Kim}
\affiliation{Centre de Physique Th\'{e}orique Ecole Polytechnique, CNRS, Universite Paris-Saclay, 91128 Palaiseau, France}
\affiliation{College de France, 11 place Marcelin Berthelot, 75005 Paris, France}
\author{A. Tamai}
\affiliation{Department of Quantum Matter Physics, University of Geneva, 24 Quai Ernest-Ansermet, 1211 Geneva 4, Switzerland}
\author{S. McKeown Walker}
\affiliation{Department of Quantum Matter Physics, University of Geneva, 24 Quai Ernest-Ansermet, 1211 Geneva 4, Switzerland}
\author{F. Y. Bruno}
\affiliation{Department of Quantum Matter Physics, University of Geneva, 24 Quai Ernest-Ansermet, 1211 Geneva 4, Switzerland}
\author{I. Cucchi}
\affiliation{Department of Quantum Matter Physics, University of Geneva, 24 Quai Ernest-Ansermet, 1211 Geneva 4, Switzerland}
\author{E. Cappelli}
\affiliation{Department of Quantum Matter Physics, University of Geneva, 24 Quai Ernest-Ansermet, 1211 Geneva 4, Switzerland}
\author{C. Besnard}
\affiliation{Department of Quantum Matter Physics, University of Geneva, 24 Quai Ernest-Ansermet, 1211 Geneva 4, Switzerland}
\author{T. K. Kim}
\affiliation{Diamond Light Source, Harwell Campus, Didcot, United Kingdom}
\author{P. Dudin}
\affiliation{Diamond Light Source, Harwell Campus, Didcot, United Kingdom}
\author{M. Hoesch} \thanks{New permanent address: Deutsches Elektronen-Synchrotron
DESY, Photon Science, Hamburg, 22607, Germany}
\affiliation{Diamond Light Source, Harwell Campus, Didcot, United Kingdom}
\author{M. J. Gutmann}
\affiliation{ISIS Neutron and Muon Source, Science and Technology Facilities Council, Rutherford Appleton Laboratory​​​, Didcot OX11 0QX, United Kingdom}
\author{A. Georges}
\affiliation{Centre de Physique Th\'{e}orique Ecole Polytechnique, CNRS, Universite Paris-Saclay, 91128 Palaiseau, France}
\affiliation{College de France, 11 place Marcelin Berthelot, 75005 Paris, France}
\affiliation{Department of Quantum Matter Physics, University of Geneva, 24 Quai Ernest-Ansermet, 1211 Geneva 4, Switzerland}
\affiliation{Center for Computational Quantum Physics, Flatiron Institute, 162 5th Avenue, New York, NY 10010, USA}
\author{R. S. Perry}
\affiliation{London Centre for Nanotechnology and UCL Centre for Materials Discovery, University College London, London WC1E 6BT, United Kingdom}
\author{F. Baumberger}
\affiliation{Department of Quantum Matter Physics, University of Geneva, 24 Quai Ernest-Ansermet, 1211 Geneva 4, Switzerland}
\affiliation{Swiss Light Source, Paul Scherrer Institut, CH-5232 Villigen PSI, Switzerland}

%\date{\today}

\maketitle

\textbf{Abstract: We report the evolution of the $k$-space electronic structure of lightly doped bulk \CRO{} with uniaxial strain.
	Using ultrathin plate-like crystals, we achieve strain levels up to $\bv{-4.1\%}$, sufficient to suppress the Mott phase and access the previously unexplored metallic state at low temperature. Angle-resolved photoemission experiments performed while tuning the uniaxial strain reveal that metallicity emerges from a marked redistribution of charge within the Ru $\bv{t_{2g}}$ shell, accompanied by a sudden collapse of the spectral weight in the lower Hubbard band and the emergence of a well defined Fermi surface which is devoid of pseudogaps. Our results highlight the profound roles of lattice energetics and of the multiorbital nature of \CRO{} in this archetypal Mott transition and open new perspectives for spectroscopic measurements.}
	
%	importance of lattice energetics in correlated metal-insulator transitions and open new perspectives for spectroscopic measurements.}

%(should estimate elastic energies and compare to electronic ones)
\vspace{1cm}

\textbf{Main text:}

Mott metal-insulator-transitions are driven by electron-electron interactions but often coincide with structural phase transitions~\cite{Imada1998}. While the latter were long believed to be a secondary response, as argued originally by N.F. Mott~\cite{Mott1949}, realistic numerical studies point to a far more important role of structural changes in stabilizing the Mott state of archetypal insulators~\cite{Pavarini2004,Gorelov2010}. This, together with recent theoretical advances, has led to renewed interest in the interplay of lattice energetics and electronic properties near Mott transitions~\cite{Park2013,Leonov2014,Han2018}.
Hydrostatic and uniaxial pressure is particularly important in the experimental study of Mott transitions and also has a profound effect on other emerging properties of quantum materials~\cite{Imada1998,Mathur1998,Schlom2007,Drozdov2015,Hicks2014,Burganov2016}. 
%Hydrostatic pressure is the tool of choice for the study of this question and plays an important role in 
However, conventional pressure cells are fundamentally incompatible with modern surface sensitive spectroscopies such as angle-resolved photoemission (ARPES). Consequently the evolution of the $k$-space electronic structure in Mott systems as they are tuned across the metal-insulator transition (MIT) has remained largely unkown. In order to overcome this limitation of ARPES, we developed an apparatus which is compatible with modern ARPES facilities and permits \textit{in-situ} quasi-continuous tuning of uniaxial strain.
%Pressure plays an important role in the study of quantum materials~\cite{Imada1998,Mathur1998,Schlom2007,Drozdov2015,Hicks2014}. Its application in angle resolved photoemission (ARPES) studies, however, has so far been limited to detwinning bulk crystals by constant uniaxial stress~\cite{Yi2011,Watson2017} and to experiments on thin films strained by lattice mismatch with the substrate~\cite{Yoo2015,Burganov2016}. 
%Our material system of choice to investigate this question is the perovskite Mott system \CRO, 
Here, we use this new capability to investigate the layered perovskite \CRO, which is of particular scientific interest as a prototypical multiband Mott insulator.
Within band theory \CRO{} is a good metal with a nearly uniform distribution of the 4 Ru $d$-electrons over the 3 $t_{2g}$ orbitals. How such a multiband metal with fractional occupation can undergo a Mott transition has been debated intensely, but the lack of data from the metallic state has prevented stringent tests of theoretical models~\cite{Anisimov2002,Liebsch2007,Gorelov2010,Zhang2017,Kubota2005,Sutter2017}. More recently the magnetic properties in the insulating state of \CRO{} have attracted much interest~\cite{Kunkemoller2015,Jain2017,Zhang2017} following proposals of a $J_{\rm{eff}}=0$ state with excitonic magnetism and an exotic doping evolution~\cite{khaliullin2013,Chaloupka2016} as well as the observation of unprecedented diamagnetism in a semimetallic phase induced by dc electric current~\cite{Sow2017}. 

%need to cite ~\cite{Yoo2015,Burganov2016}

The insulating state of \CRO{} is known to be very sensitive to pressure~\cite{Nakamura2002,Steffens2005,Miao2012,Dietl2018}, chemical substitution~\cite{Nakatsuji2000,Fukazawa2001} and even electric fields~\cite{Nakamura2013,Sow2017}.
To further increase the sensitivity of the insulating ground state of \CRO{} to strain, we have grown a series of La, Nd and Pr doped single crystals. Details of the sample growth and characterization are given in the supplementary information. Despite the slightly different rare earth ionic radii these samples behave qualitatively similarly. We thus chose to concentrate on \CPRO{} with $x=0$, 0.03, 0.04, 0.07. Consistent with a previous study on La-doped \CRO~\cite{Fukazawa2001} we find that Pr does not introduce itinerant carriers in the insulating state but suppresses the structural phase transition accompanying the metal insulator transition (MIT) from $T_{MI}\sim360$~K for $x=0$ to $\sim85$~K at the highest doping level of $x=0.07$ used in our study (see phase diagram in Fig.~1{\bf a}). Our single crystal neutron diffraction data show that the structural transition of \CPRO{} is similar to the one in pure \CRO{}. In particular, it is symmetry-preserving for all doping levels and mainly characterized by a flattening of the RuO$_{6}$ octahedra together with an elongation of the $b$-axis leading to strong orthorhombicity in the insulating phase [see supplementary information]. We will exploit this latter property to tune the MIT by uniaxial strain. Adopting the notation used for pure \CRO, we call the metallic phase with long $c$-axis and $Pbca$ space group \LPbca{} and the insulating phase with short $c$ axis \SPbca.

\begin{figure*}[!htp]
	\includegraphics[width=16cm,clip]{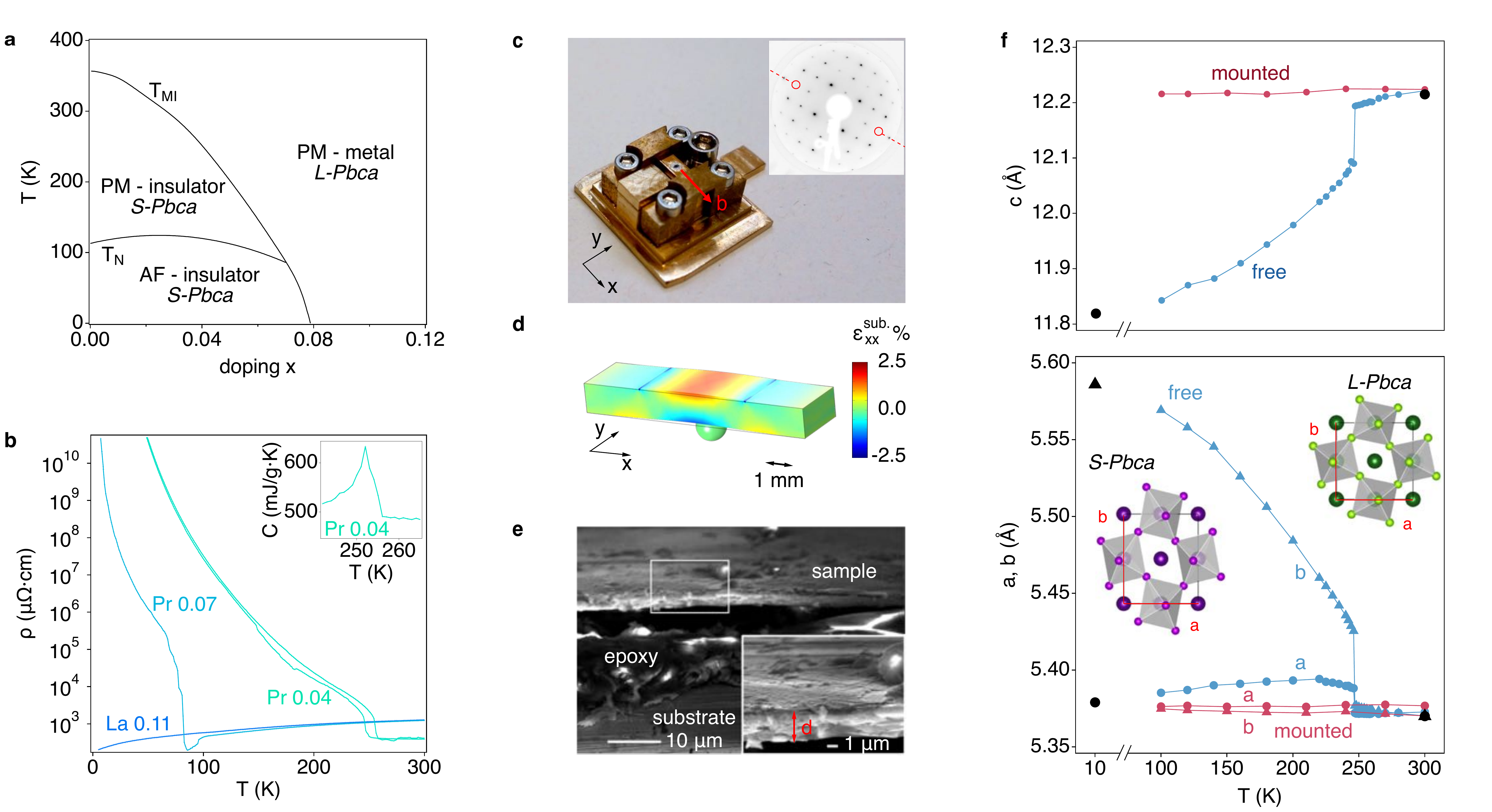} %
	\caption{{\bf Phase diagram and strain apparatus}. {\bf a} Metallic and insulating phases of \CPRO{} are separated by a first order structural phase transition from \LPbca{} to \SPbca{}. Canted antiferromagnetism is observed in all insulating samples below $\approx 110$~K.
		{\bf b} Resistivity curves for \CPRO{} ($x=0.04,0.07$) and \LCRO{} ($x=0.11$), which we use as a reference for the metallic ground state. The hysteretic behavior shown for $x=0.04$ confirms that the transition is first order. The inset shows the peak in the specific heat at the structural phase transition for $x=0.04$.
		{\bf c} Photograph of the strain apparatus. Bending the substrate along the $b$ axis drives strained \CPRO{} towards the insulating orthorhombic \SPbca{} ground state. The inset shows a LEED pattern of a strained sample, revealing the glide plane (dashed red line).
		{\bf d} Calibration of the strain apparatus using finite element analysis. The color scale encodes the tensile strain.
		{\bf e} Scanning electron micrograph of a cleaved and fully strained sample. The black region between sample and epoxy layer is due to a shadowing effect caused by the high roughness of the cut through sample and substrate.
		{\bf f} Temperature dependence of the lattice constants for $x=0.04$ measured by single crystal X-ray diffraction (XRD) before and after mounting the sample on our strain apparatus. We find that samples as thin as the one imaged in {\bf e} preserve the high-temperature \LPbca{} structure down to base temperature. Black symbols indicate lattice constants obtained by single crystal neutron diffraction at $10$~K and $300$~K.}
	\label{f1}
\end{figure*}

Our \textit{in-situ} transferable strain apparatus is shown in Fig.~1{\bf c}. It is actuated mechanically by turning a screw, which causes a lever to press a stainless steel ball from below on a 1~mm thick CuBe substrate. 
The elastic deformation of the substrate results in tensile strain $\epsilon_{xx}^{sub.}$ along the bending direction on the upper surface and a much smaller compressive strain $\epsilon_{yy}^{sub.}$ in the orthogonal direction. We calibrate $\epsilon^{sub.}$ using finite element analysis, as shown in Fig.~1{\bf d} and supplementary Fig.~2, taking into account the indent in the substrate left by the ball, which we measure at the end of each experiment.
For a maximal coupling of in-plane strain to the $c$-axis compression, which putatively drives the MIT~\cite{Gorelov2010}, we align the crystalline $b$-axis with the bending direction. Since this axis lies in a glide plane of the $Pbca$ structure, it can be identified readily in low-energy electron diffraction (LEED) patterns via the extinction of spots at certain energies (inset to Fig.~1{\bf c}).

Key to our experiment is the exploitation of the initial compressive strain exerted by the large differential thermal contraction as apparatus and sample are cooled to base temperature. Using literature data for the CuBe substrates and our neutron diffraction data for \CPRO, we calculate nominal values of $\epsilon_{xx}^{i}=-4.1\%$ $(-2.3\%)$ for Pr concentrations \mbox{$x=0.04$ (0.07)} [see supplementary information].
We directly confirm these exceptionally high strain levels for the most challenging case of a $x=0.04$ sample using X-ray diffraction on cleaved samples mounted on our strain apparatus. From the data shown in Fig.~1{\bf f} we calculate an initial compressive strain $\epsilon_{xx}^{i}=(b^{\mathrm{mounted}}-b^{\mathrm{free}})/b^{\mathrm{free}}=-3.6\%$ at 100~K, in excellent agreement with the nominal value of $-3.8\%$ at this temperature. 
These strain levels are achieved by mounting ultrathin plate-like single crystals to minimize strain relaxation. Cross-sectional electron microscopy images of our mounted and cleaved samples indicate typical thicknesses of $\sim10\;\mu$m for the epoxy layer and $2-10\;\mu$m for the single crystals (Fig.~1{\bf e}).
Having confirmed negligible relaxation at the highest strain used in our experiment, we approximate the total strain as $\epsilon^{tot}=\epsilon^{i}+\epsilon^{sub.}$, where $\epsilon^{i}$ is compressive and $\epsilon^{sub.}$ tensile.

\begin{figure*}[!ht]
	\includegraphics[width=16cm]{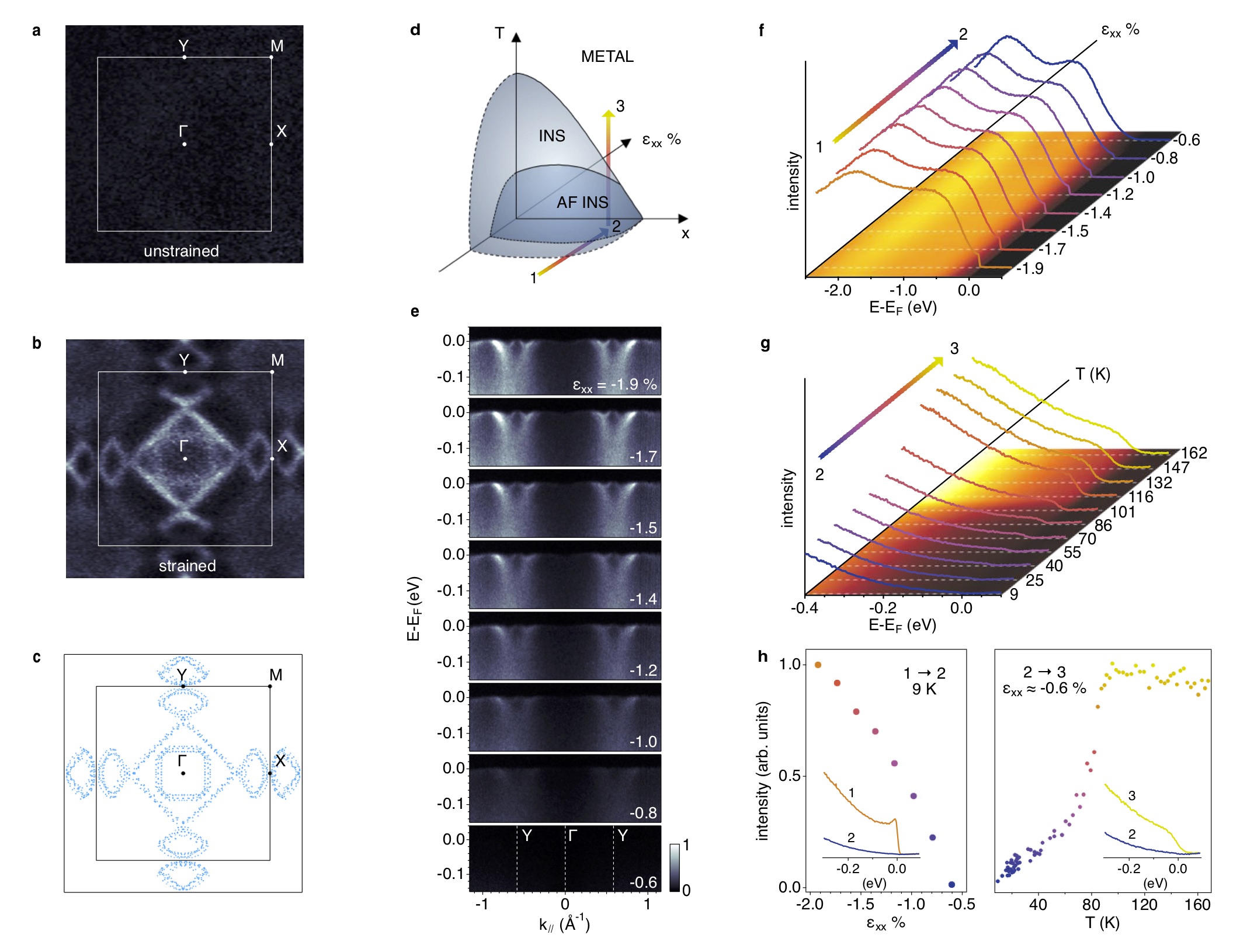} %
	\caption{{\bf Strain tuning of the MIT and nature of the low-temperature metallic state.} {\bf a,b} ARPES Fermi surface maps for a $x=0.07$ sample and a fully strained $x=0.04$ sample measured at 50~K and 8~K, respectively. The former was measured on a sufficiently thick sample to cause almost complete relaxation of the initial strain. The data were acquired using $64$~eV photons with linear horizontal polarization.
		{\bf c} Fermi surface contours extracted from the data in {\bf b}.
		{\bf d} Schematic three dimensional \textit{strain-doping-temperature} phase diagram of \CPRO.
		{\bf e} Evolution of the quasiparticle band structure at 8~K along Y$\Gamma$Y as the strain is tuned along the path $1\rightarrow2$ in {\bf a}.
		{\bf f} Angle-integrated energy distribution curves (EDCs) over the full width of the occupied Ru $t_{2g}$ states as a function of uniaxial strain.
		{\bf g} Angle-integrated EDCs as a function of temperature measured at minimum strain $\epsilon_{xx}^{i}\approx-0.6\%$ (path $2\rightarrow3$). The sample undergoes the MIT at $\sim 90$~K.
		{\bf h} Evolution of the spectral weight at the Fermi level along the path $1\rightarrow2\rightarrow3$ defined in {\bf d}.}
	\label{f2}
\end{figure*}

From Fig.~1{\bf f} it is evident that suitably mounted samples remain in the quasi-tetragonal \LPbca{} structure down to base temperature. Bending the substrate in our strain apparatus at low temperature then drives \CPRO{} towards the orthorhombic \SPbca{} ground state. 
%In Fig.~2, we demonstrate the use of strain to control the electronic state of \CPRO{} and we investigate the low-temperature metallic phase induced by compressive strain. 
The striking effect of uniaxial strain on the electronic structure of \CPRO{} is evident from the ARPES Fermi surfaces shown in Fig.~2{\bf a,b}. For an unstrained sample in the \SPbca{} phase we find negligible intensity at the Fermi level $E_F$ and no discernible structure in momentum space consistent with a gapped Mott insulating state. In a fully strained sample with \LPbca{} structure, on the other hand, a clear Fermi surface emerges, demonstrating a metallic ground state.
Intriguingly, the strain-induced metallic state differs strongly from lightly doped cuprates and iridates~\cite{Damascelli2003,DeLaTorre2015}. 
In particular, we find no anisotropy in the quasiparticle coherence and no evidence for a pseudogap along the entire Fermi surface within the precision of our experiment of $\approx2$~meV.
A cut along the $\Gamma$Y high-symmetry line (Fig.~2{\bf e}, first panel) shows well defined, strongly renormalized quasiparticle states at very low energy only indicating a delicate Fermi liquid regime. Beyond a coherence scale of $\sim30$~meV, the excitations broaden rapidly and their dispersion increases simultaneously. These high-energy states can be tracked down to $\sim-2.7$~eV and thus essentially over the full bare band width (see Fig.~3). Such a coexistence of heavy quasiparticles with 'unrenormalized' high-energy states was identified as a hallmark of Hund's metals with profound implications on magnetic susceptibility, thermal and electrical transport~\cite{Georges2012}.

Subsequently, we use our strain apparatus to tune across the Mott transition. Fig.~2{\bf e,f} shows the evolution of the near-$E_F$ electronic structure for a $x=0.07$ sample following the path $1\rightarrow2$ in the schematic phase diagram of Fig.~2{\bf d}. Clearly, we achieve a sufficient tuning range to fully recover the characteristic spectrum of insulating \CPRO{} with an exponential onset of weight. The angle-resolved data show that the Mott transition is approached by a uniform reduction of the spectral weight. Interestingly though, the quasiparticle dispersion is not affected strongly by strain. We can thus exclude that the strain-induced Mott transition is triggered by a divergence of the effective mass predicted in the Brinkmann-Rice model~\cite{Brinkmann1970}.
Raising the temperature ($2\rightarrow3$, Fig.~2{\bf g}), the insulating state undergoes another phase transition close to $T_{MI}$ of the unstrained state and we recover a metallic spectrum with significant weight at $E_F$. As shown in Fig.~2{\bf e,f}, the suppression of the spectral weight at $E_F$ during the strain-tuning is gradual. This can either indicate a second order phase transition or a phase coexistence with domains below the lateral dimension of $\approx 20\times50\:\mu$m probed by ARPES. Given the sensitivity of the electronic state of \CPRO{} to its first order structural phase transition, we consider the latter more likely. Additional evidence for phase coexistence, which was also observed in diffraction experiments on \CRO{} under hydrostatic pressure~\cite{Steffens2005}, is shown in supplementary information.

The Fermi surface of strained \CPRO{} is remarkably simple considering the large unit cell containing 4 formula units and 16 electrons in the Ru $t_{2g}$ shell. We find a square hole-like sheet centered at $\Gamma$, which encloses a smaller electron-like Fermi surface, and four small lens-shaped sheets at the $X$ and $Y$ points, respectively. The absence of exchange splitting in our experimental data indicates a paramagnetic metallic state, as it is also observed in the \LPbca{} phase of undoped \CRO{} and for highly La-doped \CRO{} with \LPbca{} structure in the ground state~\cite{Fukazawa2001}. We thus conclude that the low-temperature metallic state induced by uniaxial strain in our experiments represents the intrinsic metallic phase of \CRO{} but differs from bulk crystals under hydrostatic pressure and from epitaxially strained thin films, where ferromagnetism with an ordered moment of 0.1 $\div$ 0.3 $\mu_{B}$ is observed below $\sim 20$~K~\cite{Nakamura2002,Miao2012,Dietl2018}. 

Measuring the Fermi surface volume, we find approximate electron-hole compensation, consistent with a localized character of the Pr excess electrons. Doping dependent measurements (not shown), on the other hand indicate a measurable change of the Fermi surface volume with doping $x$, suggesting an at least partial delocalization of the dopants in the metallic \LPbca{} phase.
Due to the large rotation of the RuO$_6$ octahedra and the sizeable spin-orbit coupling, we cannot uniquely identify the orbital character on the Fermi surface from linear dichroism measurements. However, it appears plausible to interpret the extended straight sections of the experimental Fermi surface as originating predominantly from the quasi-1d $xz$,$yz$ orbitals while the curved sections of the lens-pockets as well as the circular pocket at $\Gamma$ are likely of dominant $xy$ character. Based on this interpretation, we estimate the orbital polarization in the metallic state using a simple tight-binding model fitted to the experimental Fermi surface contours. This suggest orbital occupations $n_{xz}\approx n_{yz}\approx 1.4$ and $n_{xy}\approx1.2$, and thus a strongly reduced polarization $p=n_{xy}-(n_{xz}+n_{yz})/2\approx$ 0 $\div$ 0.4 with respect to the insulating state 
characterized in previous work~\cite{Kubota2005,Gorelov2010,Sutter2017,Han2018}.
%with completely filled $xy$ orbital, half-filled out-of-plane orbitals and $p=1$~\cite{Kubota2005,Gorelov2010,Sutter2017,Han2018}. 

Further evidence of a large redistribution of spectral weight at the strain-induced MIT comes from the ARPES data over a larger energy scale. In Fig.~3{\bf a,c} we compare dispersion plots covering the full width of the $t_{2g}$ shell from two insulating and paramagnetic samples with \SPbca{} structure and $x=0, \: 0.03$ with a fully strained sample of comparable doping $x=0.04$ in the \LPbca{} structure.

\begin{figure*}[t]
	\includegraphics[width=16cm]{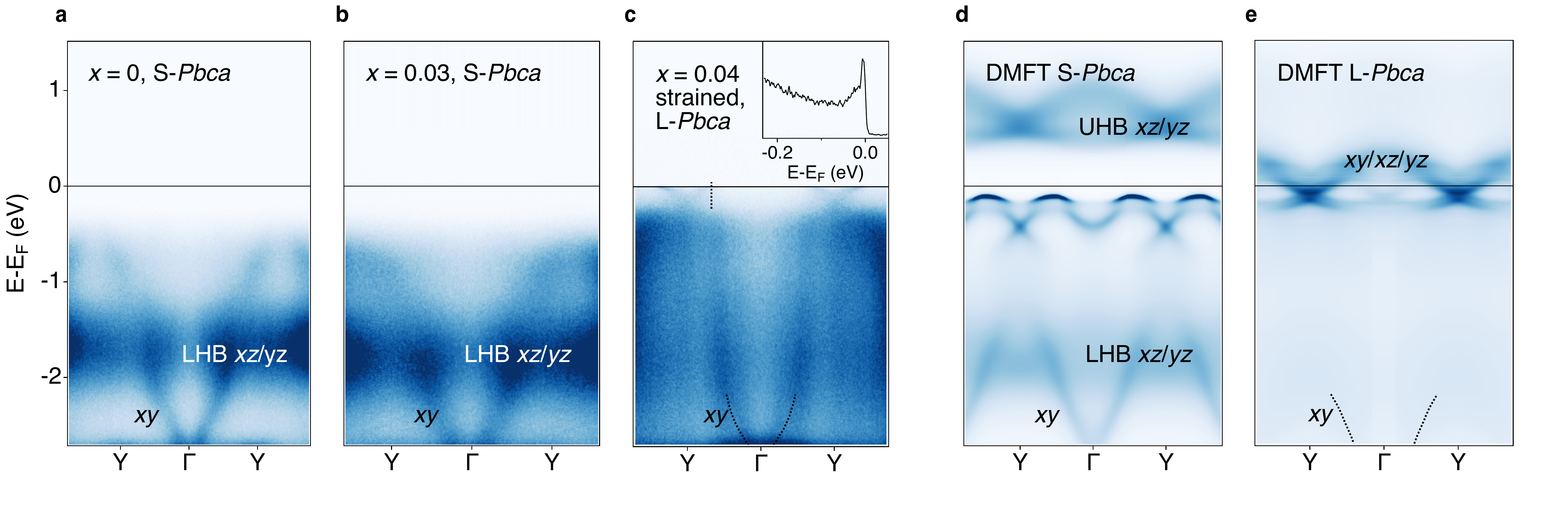} %
	\caption{{\bf Redistribution of spectral weight across the MIT.} ARPES spectral weight along $\Gamma$Y measured for different dopings and structures: {\bf a} undoped \CRO{} at 180~K (\SPbca, paramagnetic); {\bf b} $x=0.03$ at 150~K (\SPbca, paramagnetic); {\bf c} fully strained $x=0.04$ at $10$~K (\LPbca, metallic). We display a superposition of data acquired with left and right circular polarization; the inset in {\bf c} shows a clear quasi-particle peak at k=k$_{F}$ (dotted line); {\bf d,e} DMFT calculation along the same cut for undoped \CRO{} in the \SPbca{} and \LPbca{} structure at T=232~K and 390~K, respectively. The bottom of the $xy$ band, which becomes visible upon enhancing the contrast in the DMFT calculation of the metallic state is indicated by a dotted line in {\bf c,e}.}
	\label{f3}
\end{figure*}

The first important conclusion from this data is that light Pr-doping alone causes minor changes in the electronic structure only. Its main effect is a small shift of the chemical potential. Importantly though, $E_F$ remains in the correlated gap, consistent with the highly insulating nature of $x=0.03$ crystals in resistivity measurements. 
We therefore conclude that the additional valence electron of Pr stays fully localized in the insulating \SPbca{} structure, possibly due to a Mott transition in the impurity band, as it was proposed for lightly La-doped Sr$_2$IrO$_4$~\cite{Battisti2017}.
The main spectroscopic features in the insulating phase are a dispersive state with $\approx2$~eV bandwidth and an intense non-dispersive peak at -1.7~eV. With reference to our dynamical mean field theory (DMFT) calculations (Fig.~3{\bf d}) and consistent with Ref.~\cite{Sutter2017}, we identify these features with the fully occupied $xy$ orbital and the lower Hubbard band (LHB) of $xz/yz$ character, respectively. This confirms a basic electronic configuration in the insulating \SPbca{} phase with fully occupied {\xy} orbital and half-filled {\xz}/{\yz} bands split into lower and upper Hubbard band, as proposed on the basis of DMFT calculations~\cite{Liebsch2007,Gorelov2010}. 

Straining a lightly doped \CPRO{} sample, we observe a substantial redistribution of spectral weight. Most notably, the intensity in the LHB collapses suddenly across the MIT and coherent quasiparticle states appear at the chemical potential (Fig.~3{\bf c}). Both of these effects are reproduced by our DMFT calculations shown in Fig.~3{\bf e}. Interestingly though, the sudden collapse of the LHB is in stark contrast to lightly doped cuprates, where metallicity emerges from a gradual transfer of spectral weight from the LHB to coherent quasiparticle states~\cite{Shen2004}. 
We interpret this as a generic manifestation of the additional orbital degree of freedom of multiband Mott insulators. In effective single band systems, such as the cuprates or iridates, the orbital occupancy can only change by the small number of doped carriers, resulting in dominantly half-filled sites retaining a strong memory of the Mott phase.
In \CPRO, on the other hand, our data show a discontinuous change of the $n_{xz/yz}$ occupancy from 1 to $\approx1.4$ in spite of the only light doping because of substantial interorbital charge transfer across the MIT. The large deviation from half filling causes a sudden collapse of the LHB and renders electronic energies comparable to lattice energies under strain, resulting in a strongly first order nature of the Mott transition.

Our results show that tuning uniaxial strain in ARPES experiments is a promising new method to study phase transitions or, more generally, structure -- property relations of quantum materials.
%Our results show that uniaxial strain is an effective way to tune phase transitions in correlated electron systems with much potential to reveal new insight into structure -- property relations. 
Potential applications of our method range from tuning magnetism, to topological phase transitions, two-dimensional van der Waals materials and unconventional superconductors showing large responses to strain, such as \SRO~\cite{Hicks2014}.
\vspace{1cm}

\textbf{Methods:}

Single crystals of \CPRO{} were grown through the floating zone (FZ) technique using a Crystal System Corporation FZ-T-10000-H-VI-VPO-I-HR-PC four mirror optical furnace. 
Samples were grown in $90\%$ oxygen pressure, and the initial Ru concentration in the polycrystalline rods was about $20 \%$ higher than the nominal value to compensate for evaporation during the growth. 
The bulk properties were thoroughly characterized by resistivity, specific heat, magnetization measurements, and single crystal neutron diffraction at the ISIS spallation neutron source~\cite{Keen2006}. 
Doping levels were measured by energy and wavelength dispersive X-ray spectroscopy (EDX/WDX) and were found to be systematically lower by $20 \%$ to $30 \%$ than in the polycrystalline growth rod.
Angle-resolved photoemission spectroscopy (ARPES) experiments were performed at the I05 Beamline of the Diamond Light Source~\cite{doi:10.1063/1.4973562}. The presented data were acquired with linearly and circularly polarized light at $64$~eV photon energy and an overall resolution of $\approx12$~meV / 0.015~\AA$^{-1}$.\\

\begin{acknowledgments}
	We thank E. Giannini, D. McMorrow, D. Pincini, D. Jaccard, C. Renner, M. Spera,  V. Pasquier  for discussions, J. Teyssier for Raman experiments on \CPRO, M. Spera for help with the numerical strain simulations and R. Pellet for machining and aligning the strain apparatus. 
%Work in London was supported by the EPSRC grant EP/N034694/1.	
The experimental work was supported by the Swiss National Science Foundation (200021-153405 and 200020-165791). Theoretical work was supported by the European Research Council grant ERC-319286-QMAC, the NCCR MARVEL of the SNSF and the Simons Foundation (Flatiron Institute). Crystal growth and characterization at UCL was supported by the EPSRC grant EP/N034694/1.
We gratefully acknowledge the Science and Technology Facilities Council (STFC) for access to neutron beamtime at ISIS, and also for the support of sample preparation at the UCL crystal growth laboratory. We would like to thank G. Stenning for help on the Smartlab XRD and Quantum Design MPMS instruments in the Materials Characterisation Laboratory at the ISIS Neutron and Muon Source.
We acknowledge Diamond Light Source for time on beamline I05 under proposal SI17381.
\end{acknowledgments}

\bibliographystyle{naturemag}
\bibliography{CRO_biblio.bib}
\end{document}